\documentclass{aastex}          %% The manuscript based on AASTeX v5.x
\usepackage{spr-astr-addons}    %% mimicing ASTR journal style
\begin{document}
%% Article title
%
\title{The Formation and Evolution of Massive Stellar Clusters in IC~4662}

%% Running heads
\shorttitle{SSCs in IC~4662}
\shortauthors{Gilbert \& Vacca}

%% Author and Affilations
\author{Andrea M. Gilbert\altaffilmark{1}} 
\affil{The Aerospace Corporation}
\and 
\author{William D. Vacca}
\affil{USRA-SOFIA}
\email{andrea.m.gilbert@aero.org} %% non-output

%% Alternate Affilations
\altaffiltext{1}{Lawrence Livermore National Laboratory}

%% Abstract
\begin{abstract}
We present a multiwavelength study of the formation of massive stellar
clusters, their emergence from cocoons of gas and dust, and their
feedback on surrounding matter.  Using data that span from radio to
optical wavelengths, including Spitzer and Hubble ACS observations, we
examine the population of young star clusters in the central starburst region
of the irregular Wolf-Rayet galaxy IC~4662.  We model the radio-to-IR
spectral energy distributions of embedded clusters to determine the properties of their
HII regions and dust cocoons (sizes, masses, densities, temperatures), and use near-IR and optical data with
mid-IR spectroscopy to constrain the properties of the embedded clusters themselves (mass, age,
extinction, excitation, abundance).  
The two massive star-formation regions in IC~4662 are excited by stellar populations with ages of $\sim 4$ Ma and masses of $\sim 3 \times 10^5$ M$_\odot$ (assuming a Kroupa IMF).  They have high excitation and sub-solar abundances, and they may actually be comprised of several massive clusters rather than the single monolithic massive compact objects known as Super Star Clusters (SSCs).  Mid-IR spectra reveal that these clusters have very high extinctions, A$_{\rm V} \sim 20-25$ mag, and that the dust in IC~4662 is well-mixed with the emitting gas, not in a foreground screen.

\end{abstract}

%% Keywords
%\keywords{}

\section{Introduction}
\label{intro}

Massive stars in starbursts are thought to form predominantly in Super Star Clusters (SSCs), which are massive, compact, young stellar clusters that are candidates for present-day analogues of young globular clusters because of their similar sizes and inferred masses \citep[e.g.][]{whitmore93,schweizer96}.  Although the great majority of SSCs could be short-lived, destined to dissolve within ~$\sim10$ megayears (Ma) and merge into the field star population \citep{fall05,mengel05}, it is clear that their massive stars dominate feedback in starbursts.  They return enriched matter  and energy to the interstellar medium (ISM) via stellar winds and supernovae, and their hard radiation fields heat and excite gas and destroy molecules and dust grains.  A high surface density of star formation produces galactic-scale superwinds that can expel the ISM from a galaxy \citep[e.g.][]{heckman00}.

SSCs are probably formed within giant molecular cloud complexes, so observations that can probe embedded dusty regions are required to detect the youngest SSCs.  
By the time SSCs have blown away enough of their natal material to be detected at visible wavelengths, they are usually at least several million years old, so their most massive individual stars are expected to have already evolved off of the main sequence.  At this stage it is difficult to infer much about SSC formation environments, because the massive stars have significantly altered them.   
Infrared and radio wavelengths are best suited to studying the youngest SSCs, and radio interferometry has revealed a population of heavily obscured, compact thermal radio sources with inferred nebular densities n$_{\rm e} > 10^3$ cm$^{-3}$ and radii of a few parsecs
that appear to be powered by 100s-1000s of massive stars.  These sources were dubbed ultra-dense HIIRs \citep[UDHIIs,][]{kobulnicky99} in analogy with Galactic ultra-compact HIIRs \citep[UCHIIs, e.g.][]{Wood89}, which are excited by one or a few massive stars on subparsec scales, with  n$_{\rm e} > 10^4$ cm$^{-3}$. 

UDHIIs are thought to be young, embedded SSCs that have not yet disrupted and expelled the majority of their natal and surrounding material.  Their hard intrinsic emission
is absorbed and reprocessed by surrounding dense gas and dust to emerge as a thermal blackbody component in the mid-IR, where it peaks at $\sim 60-100\ \mu$m depending on the temperature and distribution of the dust \citep[e.g.][]{vacca02}. 
 In the radio regime, UDHIIs are identified by a thermal bremsstrahlung spectrum that is self-absorbed below a few cm \citep[e.g.][]{johnson03}.   
These features dominate the spectral energy distribution (SED) of UDHIIs and can be used along with any escaping visible light emission to constrain the temperature, density, and geometry of the gas and dust excited by the SSC, as well as the age and stellar content of the cluster itself \citep[e.g.][]{martinhernandez05,martinhernandez06}.  Spectroscopic measurements provide further constraints on these parameters as well as the excitation and composition of the emitting material \citep[e.g.][]{verma03}.

\begin{figure}[t]
\begin{center}
\includegraphics[width=2.5in]{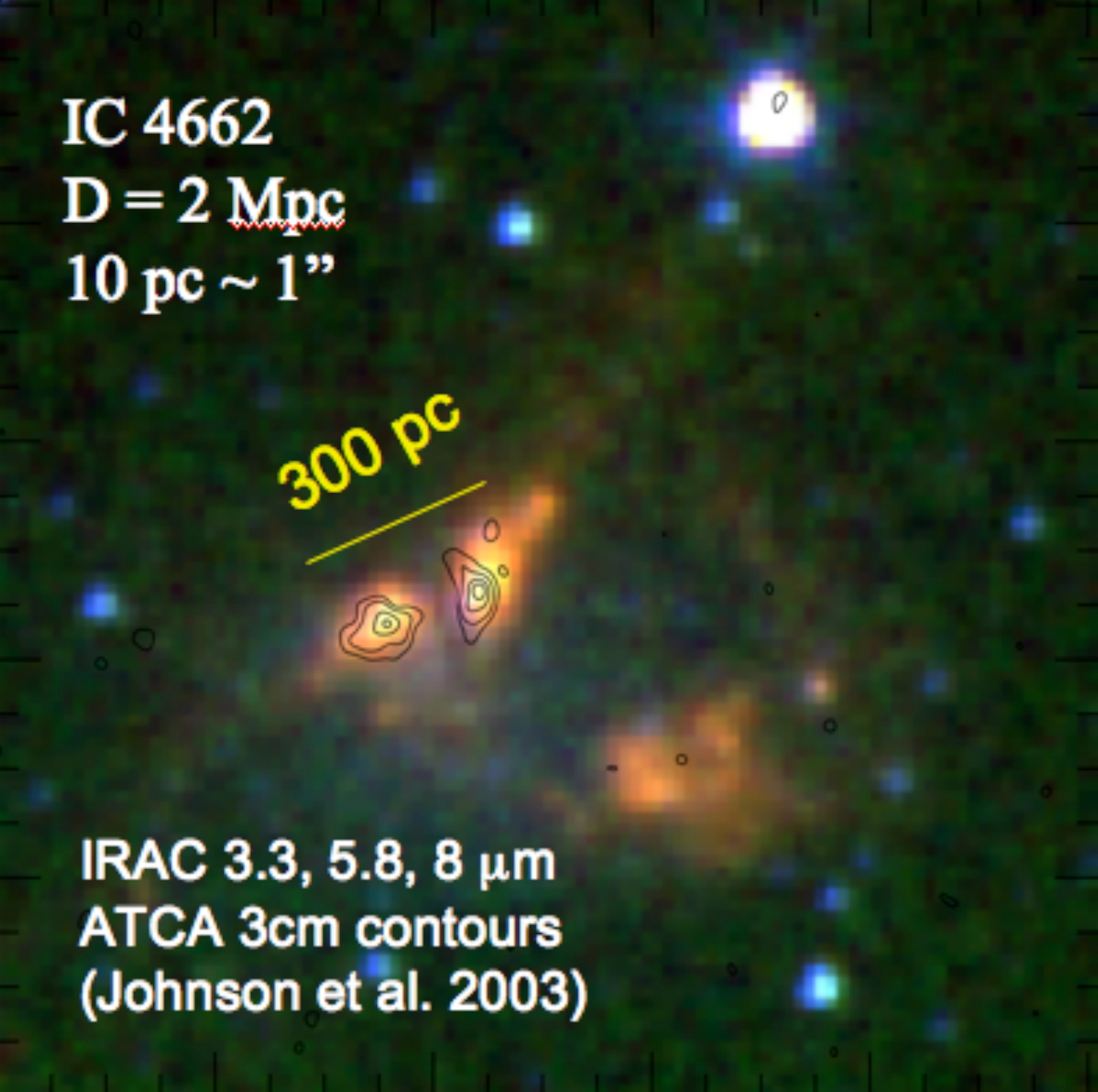}
\caption{ IRAC image of IC~4662 (with red, green, blue in 3.3, 5.8, 8 $\mu$m bands respectively) features three dusty regions that are coincident with the optically identified star-formation regions A2, A1, and B (from left to right) of \citet{heydarimalayeri90}.  Bright thermal radio sources are shown in ATCA 3 cm contours  \citep[levels start at 3$\sigma$, steps are 3$\sigma$,][]{johnson03}}
\label{fig:irac} 
\end{center}
\end{figure}

\section{Spitzer Imaging of Embedded Massive Star-Formation Regions in IC 4662}
\label{imaging}

IC~4662 is a nearby (d $\approx 2$ Mpc, 1$\arcsec \sim 10$ pc),
 isolated dwarf galaxy that contains two bright Wolf-Rayet (WR) star-forming regions that are extended over $\sim 50$ pc, and very little molecular gas and dust \citep[][]{heydarimalayeri90}.  Radio observations at 2 and 6 cm revealed that compact thermal sources containing up to a few hundred massive stars are associated with the WR star-forming regions, but they are not centered on the brightest optical clusters within them \citep{johnson03}.  We obtained Spitzer IRAC and MIPS imaging and IRS high-resolution spectroscopy of these regions in order to explore the mid-IR properties of the stars, gas, and dust in these regions.  We also obtained Hubble Space Telescope (HST) ACS imaging to study the optical sources in these regions at high resolution, but we focus on Spitzer data in this contribution and treat the starburst regions as unresolved.

\begin{figure}[t]
\begin{center}
\includegraphics[width=2.5in]{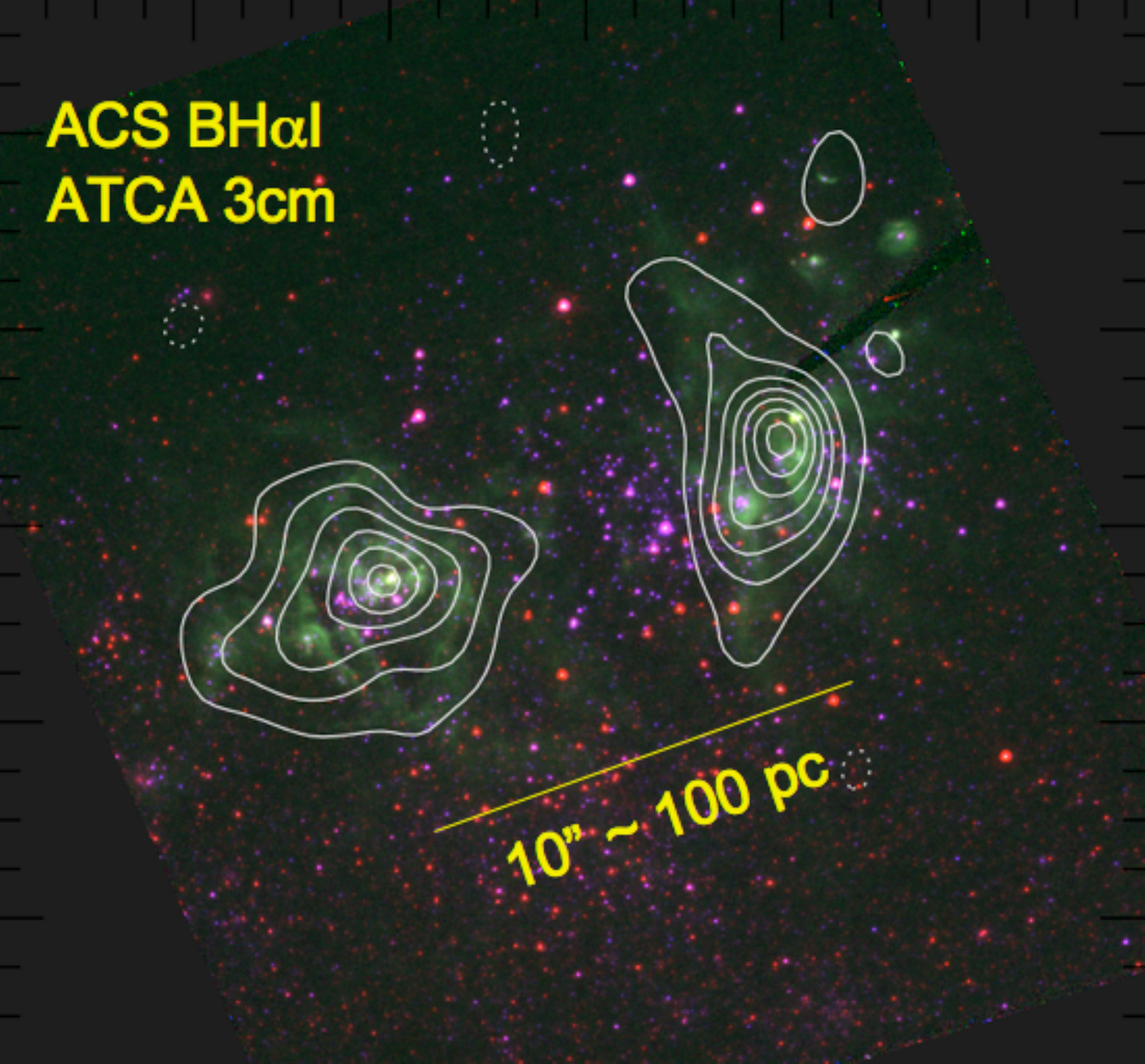}
\includegraphics[width=2.5in]{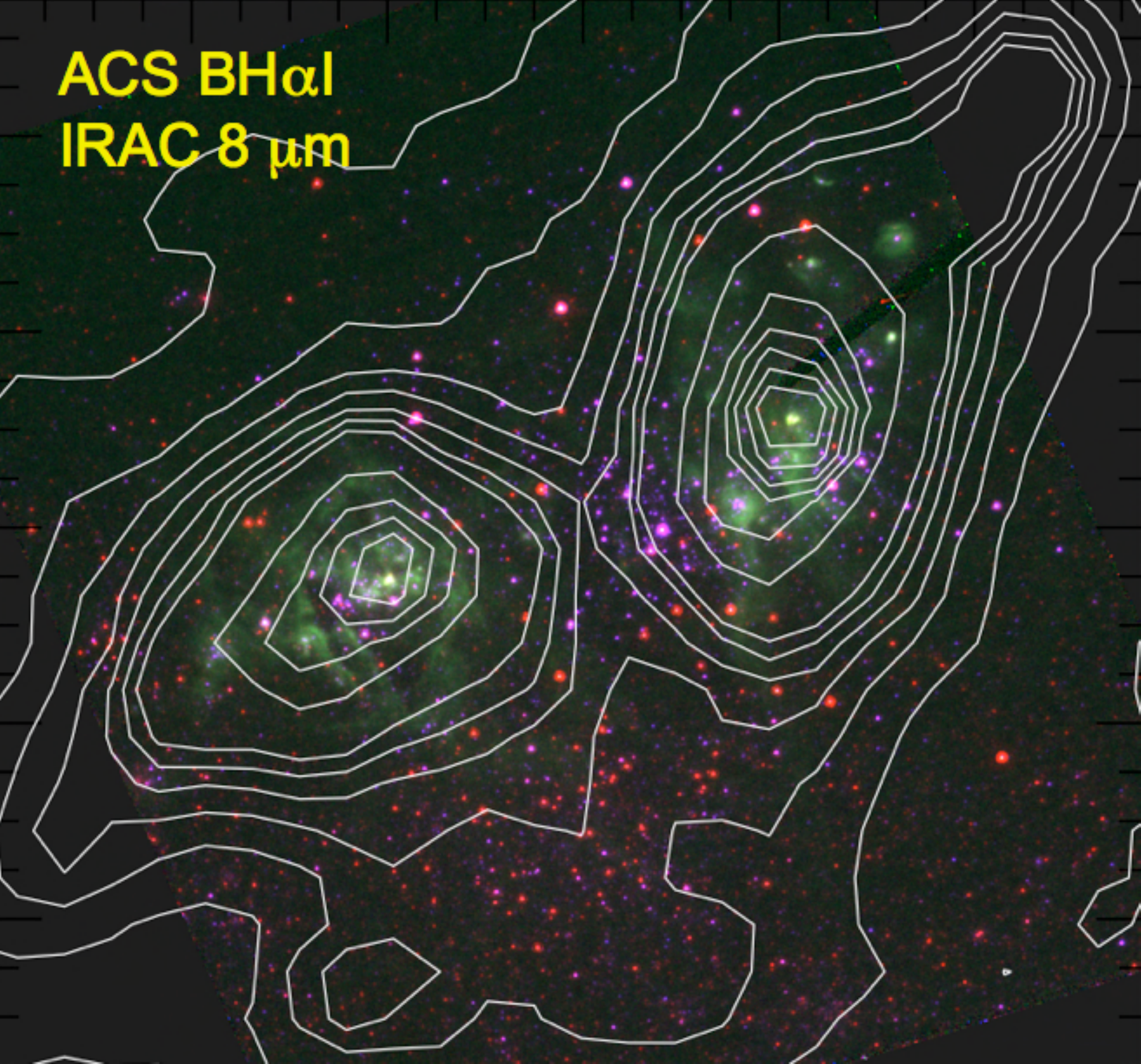}
\caption{Top: ACS image of  IC~4662  (red, green, blue in F435W, F658N, F814W, respectively) shows regions A2 and A1 overlaid with 3 cm radio contours.  H$\alpha$ and radio continuum emission have similar overall morphologies, but H$\alpha$ has filaments, sharp edges and apparently empty regions. Bottom:  Overlay of 8 $\mu$m contours shows that warm dust emission peaks where H$\alpha$ peaks}
\label{fig:bhai_3cm}
\end{center}
\end{figure}

A three-color image of IC~4662, constructed from the IRAC 3.3, 5.8, and 8-$\mu$m bands, is shown in Figure~\ref{fig:irac}, along with an overlay of the 3 cm radio contours from \citet{johnson03}.  The galaxy has three bright extended regions in the IRAC bands, two of which correspond to the radio and optical regions A1 and A2, and a third (region B, not included in the HST field of view) that is offset to the southwest and fainter in all bands.  Figure~\ref{fig:bhai_3cm} compares both the radio and IRAC 8 $\mu$m contours with the optical emission on an ACS three-color image (in filters F435W, F658N, F814W).  The radio contours follow the extended H$\alpha$ morphology, and the extended 8 $\mu$m emission, which is a tracer of PAHs or hot dust depending upon metallicity \citep[e.g.][]{engelbracht05,wu06}, peaks at the H$\alpha$ peaks.  Regions A1 and A2 are extended on scales of $\sim 50\ {\rm pc} = 5"$,  but they are complex and may contain sources that are too compact to be resolved by Spitzer's smallest pixels, which are 1.2" for IRAC.

\begin{figure}[t]
\includegraphics[width=\columnwidth]{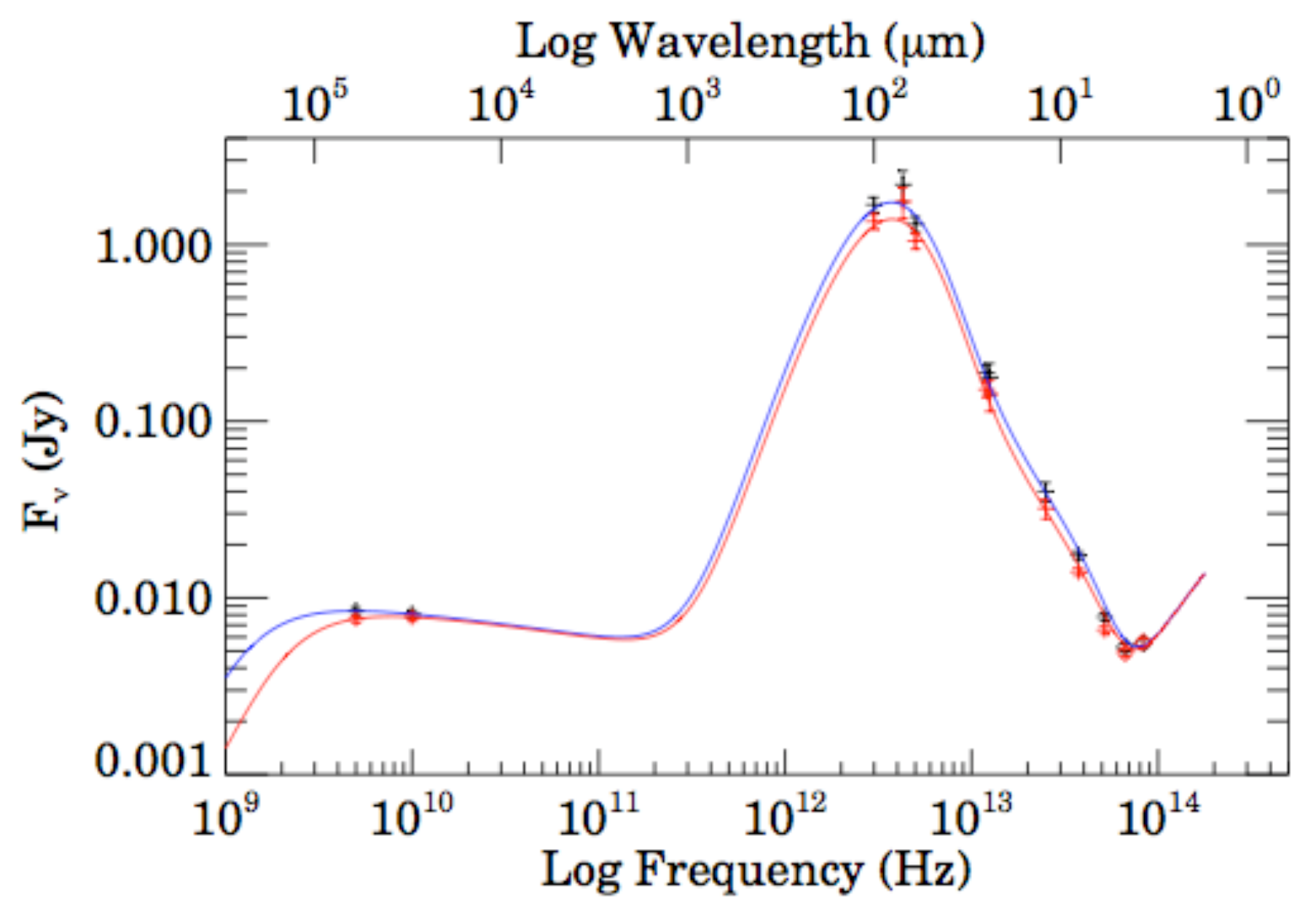}
\caption{ IC~4662 regions A1 (black, blue) and A2 (red) SEDs and fits to a model of an SSC embedded in a UDHII embedded in a dust cocoon, plus a stellar SED at $\lambda < 5\ \mu$m}
\label{fig:seds} 
\end{figure}

We performed photometry on regions A1 and A2 in the Spitzer pipeline-reduced IRAC and MIPS data.  The apertures were matched  across all Spitzer bands at a radius of $\sim 7"$, for which we also recomputed ATCA radio fluxes for consistency.  The IR background was subtracted using Spitzer SPOT predictions.  In order to incorporate the lower-resolution IRAS measurements \citep{beichman88} into our analysis we scaled the IRAS fluxes so that the 25 $\mu$m flux matches a power-law fit to the Spitzer 24 and 70 $\mu$m data interpolated at 25 $\mu$m.   This yields three additional independent estimates of fluxes of A1 and A2 at 12, 60, and 100 $\mu$m.  The resulting SEDs extend from 3 cm to 3 $\mu$m and feature a radio free-free component  longward of about 150 $\mu$m, a warm dust  blackbody peaking at around 100 $\mu$m, and the onset of stellar photospheric emission appears in the 3 $\mu$m point (Figure~\ref{fig:seds}).

\begin{figure}[t]
\begin{center}
\includegraphics[width=2.5in]{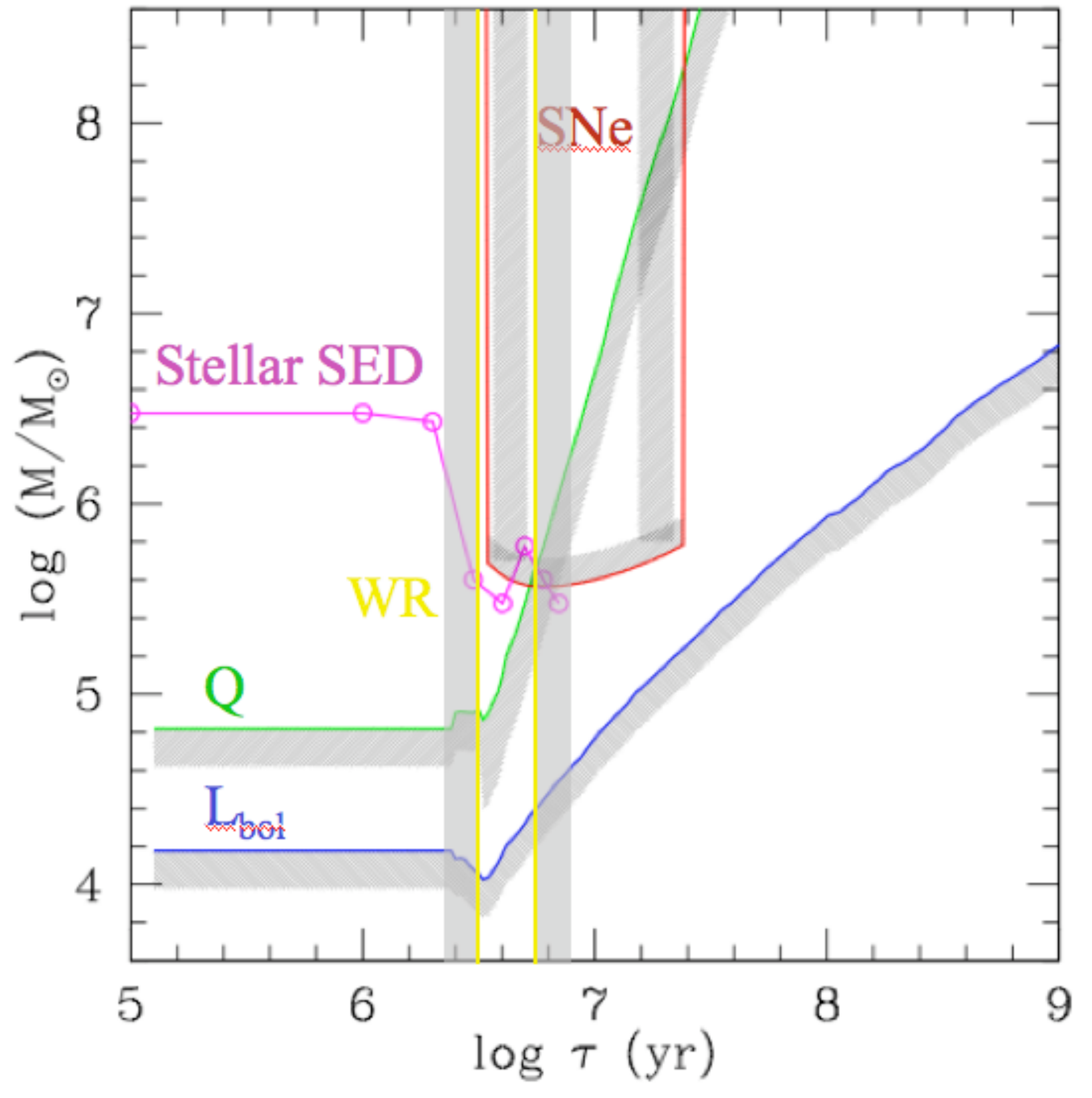}
\caption{Constraints on age $\tau$ and mass M for the exciting clusters in IC~4662 regions A1 and A2: radio flux requires ionizing photon rate Q  $ \geq 10^{51.4}$ s$^{-1}$ (above green line) and bolometric luminosity L$_{\rm bol} \geq 10^7$ L$_\odot$ (above blue line); lack of  observed 3 cm non-thermal emission implies that the clusters are outside of the SNe box (to left/right of or below red line); fit to the stellar part of the SED ($\lambda < 5 \mu$m) requires  M $\sim 3-30 \times 10^5$ M$_\odot$ (at ages on the magenta line); detection of optical and WR lines implies ages of $3-5$ Myr (between the yellow lines).  In summary, the exciting clusters of A1 and A2 must have ages of about 4 Ma and masses of about $3 \times 10^5$ M$_\odot$}
\label{fig:agemass} 
\end{center}
\end{figure}

\section{Modeling the SEDs of UDHIIs}
\label{seds}

In order to interpret the measured SEDs, we begin by considering a simple model\footnote{We will consider a more complete and complex model such as DUSTY \citep{nenkova00} in future work, as we conclude in \S~\ref{dust} that IC~4662 requires a mixed model for its dust  rather than a foreground screen.} of a star cluster embedded in an HII region which is embedded in a dust cocoon, following \citet{vacca02} and motivated by the UCHII models of e.g. \citet{churchwell90}.  The SED produced by this system is the sum of an optically thick, radio free-free component (assumed to be a constant-density, dust-free HIIR at a temperature of $10^4$ K extending from the center to an inner radius) and an optically thin dust shell component (extending from the inner to an outer radius with a constant density of silicate grains and a power-law temperature profile).  We fit the SEDs from 3 cm to 5 $\mu$m of A1 (A2) to this model and arrive at HIIR parameters: half-light radii of 4 (2.4) pc, n$_{\rm e} = 1.2\ (2.6) \times 10^3$ cm$^{-3}$, and gas masses of about $2 \times 10^5$ M$_\odot$.  The dust shells have temperatures at the inner radius of $465-485$ K and masses of about $10^3$ M$_\odot$.  The low model gas and dust masses relative to other UDHIIs like those in He~$2-10$ \citep{vacca02}
suggest that the exciting clusters A1 and A2 are more evolved, and have destroyed or ejected more of their surrounding matter in the process of emerging from their cocoons.  

Figure~\ref{fig:seds} shows that the HIIR+dust cocoon model fits the data points well, with the exception of the near-IR filters that are dominated by stellar continuum emission and may contain PAH emission at 3.3$\mu$m.  The radio data provide several constraints on the age and mass of the ionizing stellar clusters within regions A1 and A2 that are illustrated in Figure~\ref{fig:agemass}.  From the radio continuum fluxes  we derive a Lyman continuum photon rate Q $\approx 2-3 \times 10^{51}$ s$^{-1}$, which requires at least a certain mass in massive stars and hence an overall cluster mass for a full initial mass function as a function of cluster age $\tau$ \citep[we adopt a Kroupa IMF,][]{kroupa01}.  Clusters in the M$-\tau$ space above the green line in Figure~\ref{fig:agemass} have enough massive stars to explain the observed Q values even if the radio fluxes are underestimates due to self-absorption.  The blue line shows a weaker constraint following the same argument for the bolometric luminosity L$_{\rm bol}$ of the exciting cluster.  The lack of non-thermal emission at 3 cm implies no observable supernova (SN) activity, which excludes the region in M$-\tau$ that is inside the red box.  The detection of He~{\sc ii} 468.6 nm line emission \citep{heydarimalayeri90} further restricts the permitted age range for the exciting sources to lie between 3 and 5 Ma, as shown by the yellow lines.  Finally, a fit to the 3.3 and 4.5 $\mu$m points in the SEDs of \citet{bruzual03} models for a range of cluster ages and masses yields $\tau$ and M values that fall on the magenta points, so the allowed cluster parameters are  very tightly constrained at $\tau \approx 4$ Ma and M $\approx 3 \times 10^5$ M$_\odot$.  Thus we have added a stellar component to our SED model from the appropriate \citet{bruzual03} model in Figure~\ref{fig:agemass}.

\begin{figure}[t]
\includegraphics[width=1.6in]{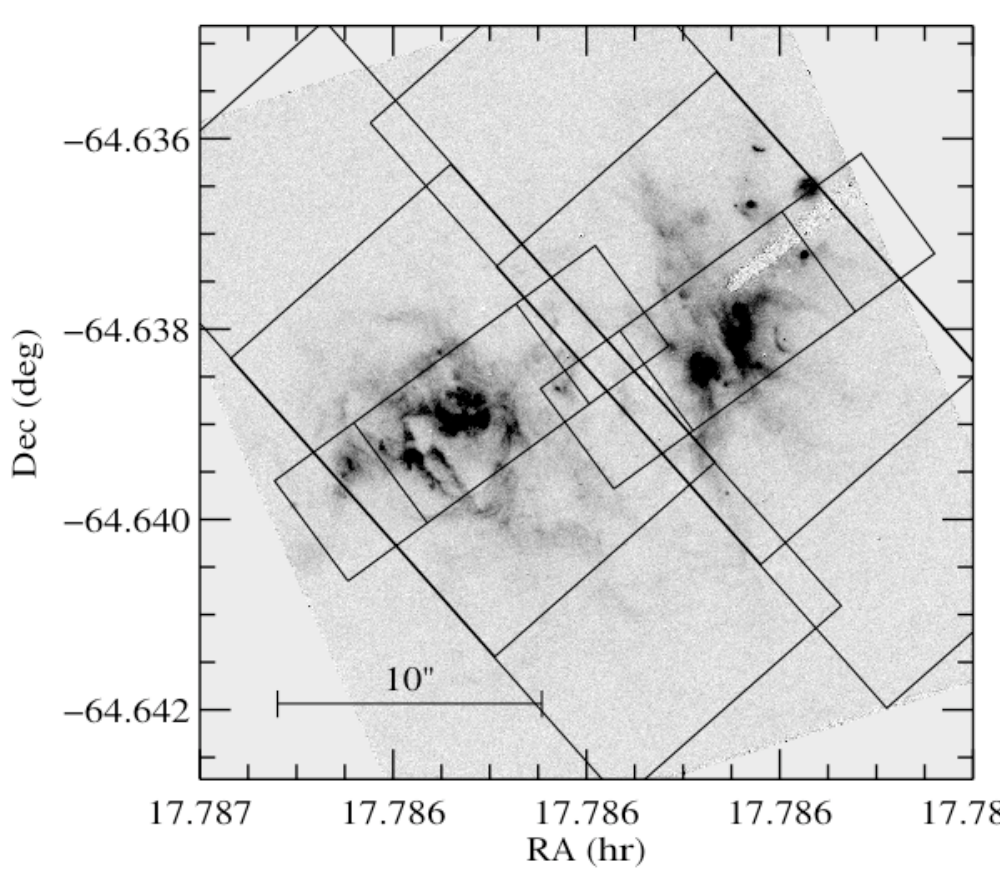}\includegraphics[width=1.6in]{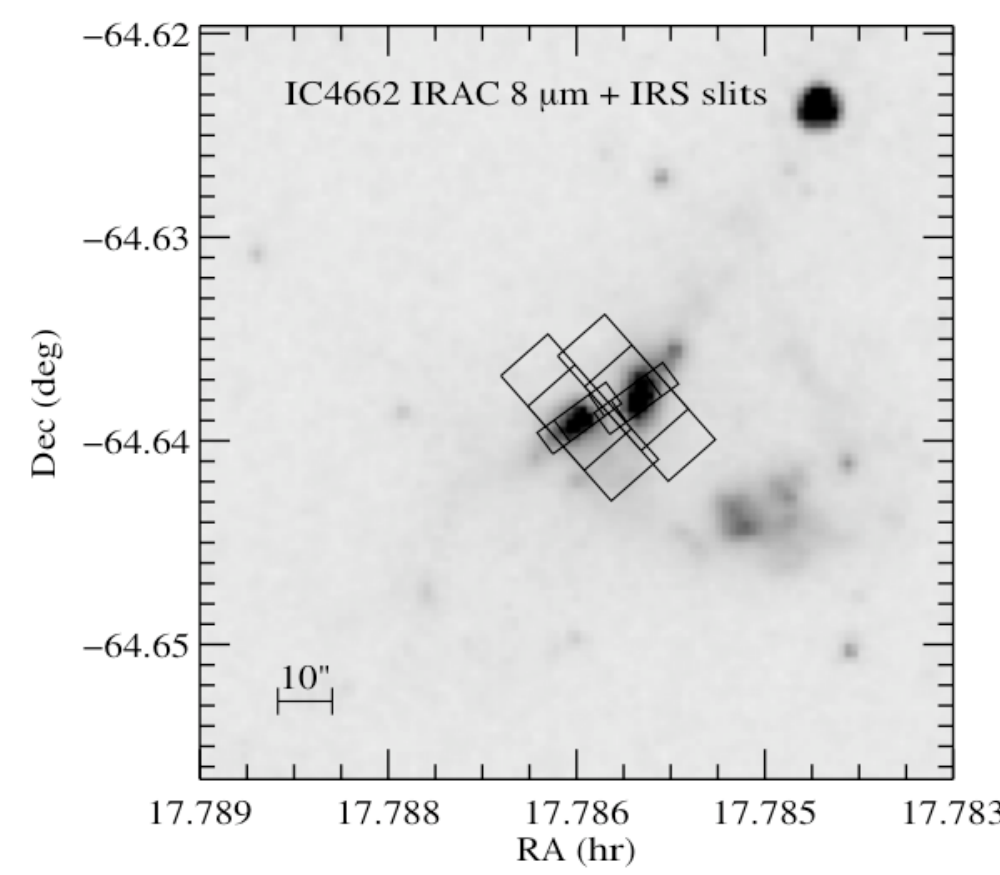}
\caption{IRS slit nod positions overlaid on IC~4662 regions A1 and A2, shown on ACS F658N (left) and IRAC 8 $\mu$m (right).  The smaller SH slits are perpendicular to the larger LH ones}
\label{fig:slits} 
\end{figure}

\begin{figure}[t]
\includegraphics[width=\columnwidth]{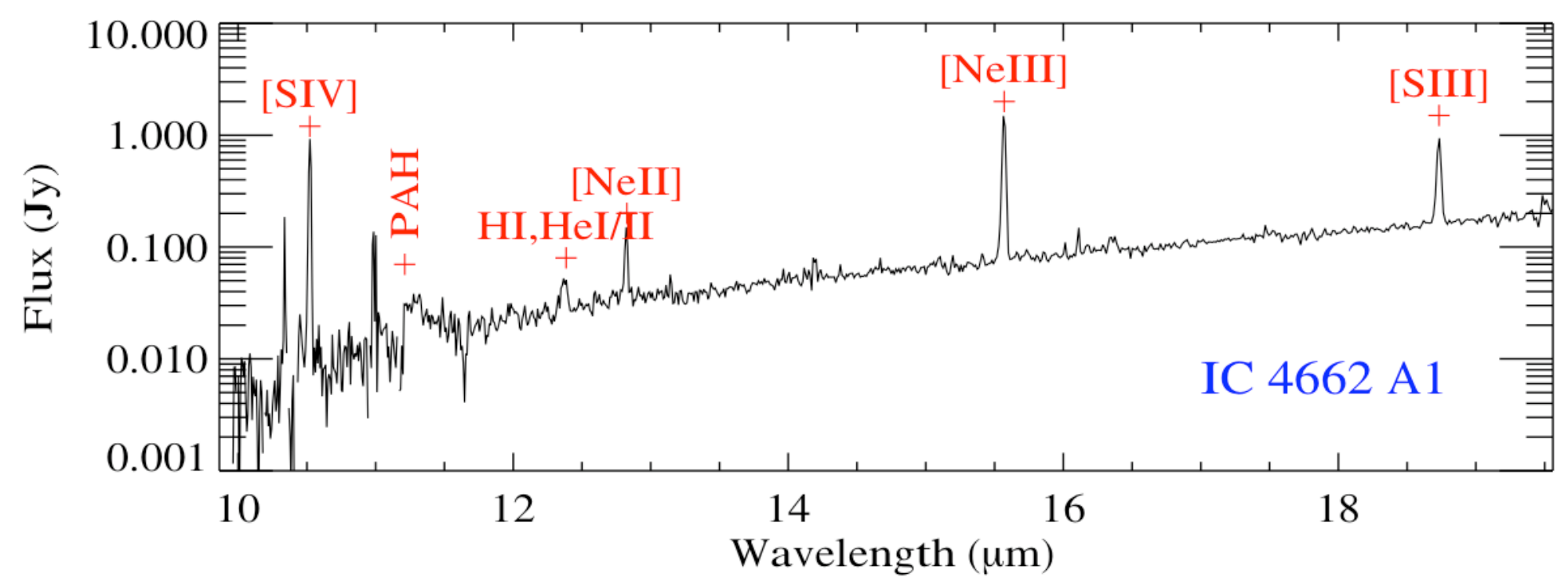}
\includegraphics[width=\columnwidth]{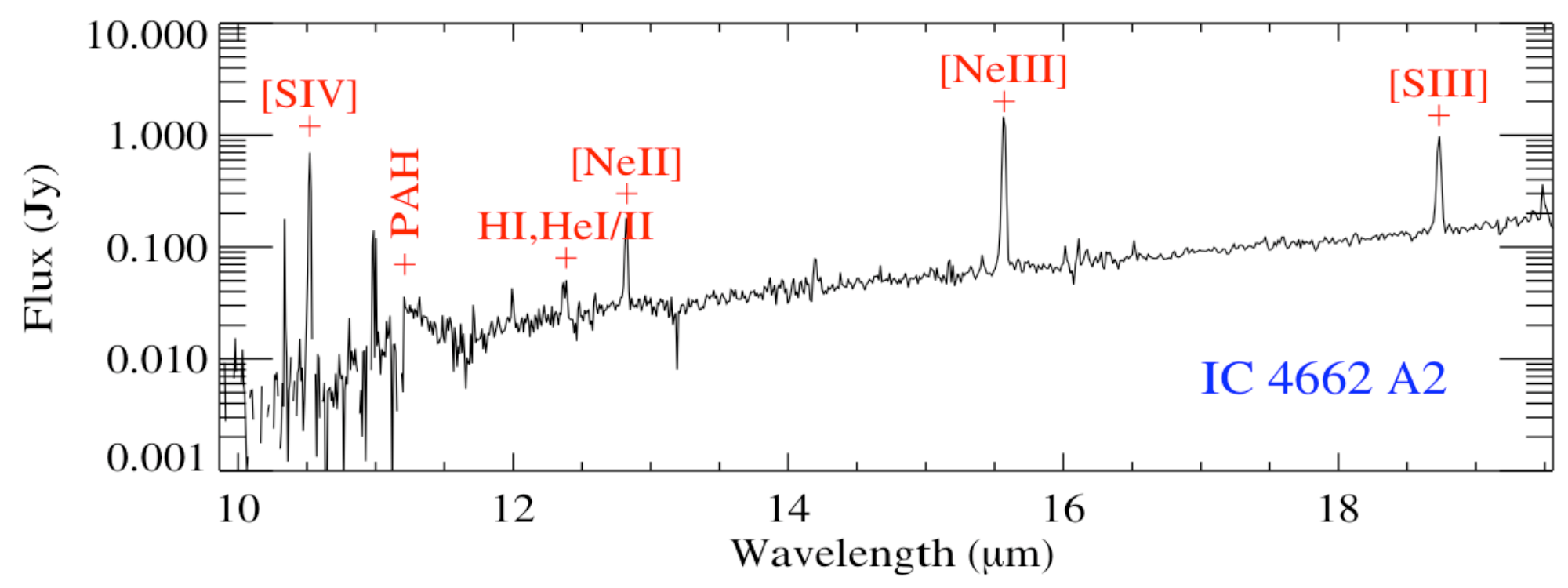}
\caption{IRS SH spectra of IC~4662 regions A1 and A2}
\label{fig:spec_sh} 
\end{figure}

\begin{figure}[h]
\includegraphics[width=\columnwidth]{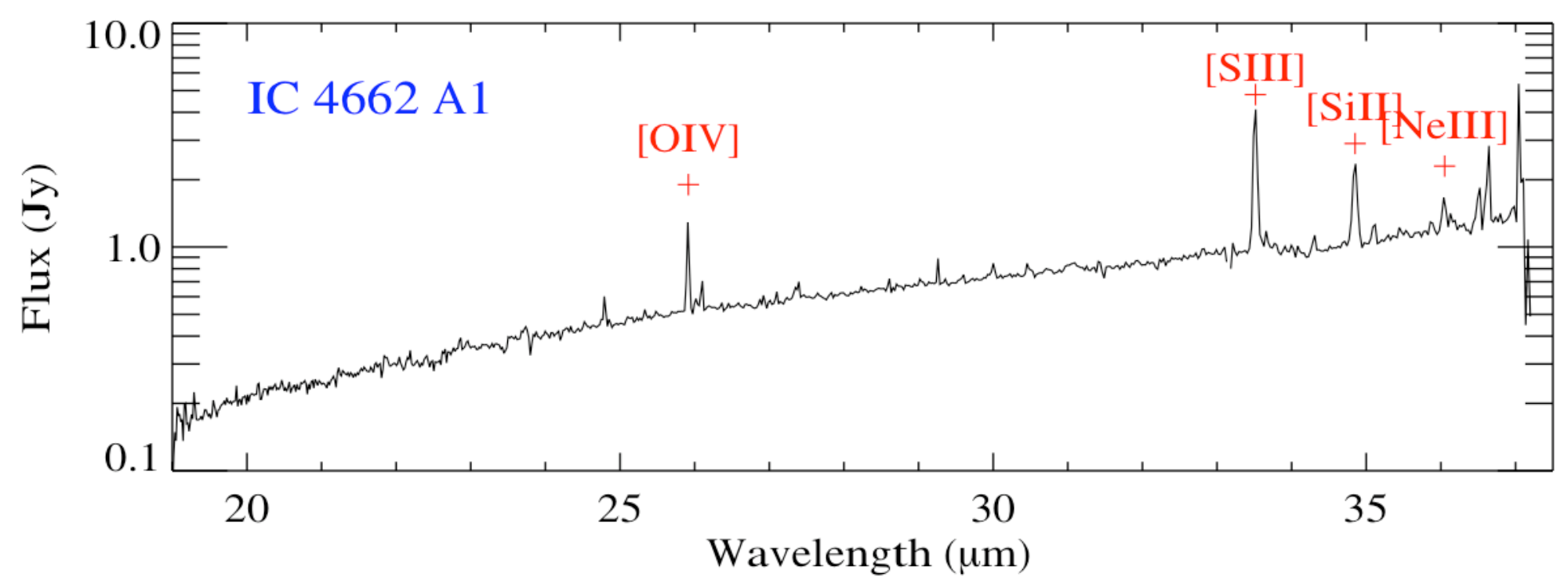}
\includegraphics[width=\columnwidth]{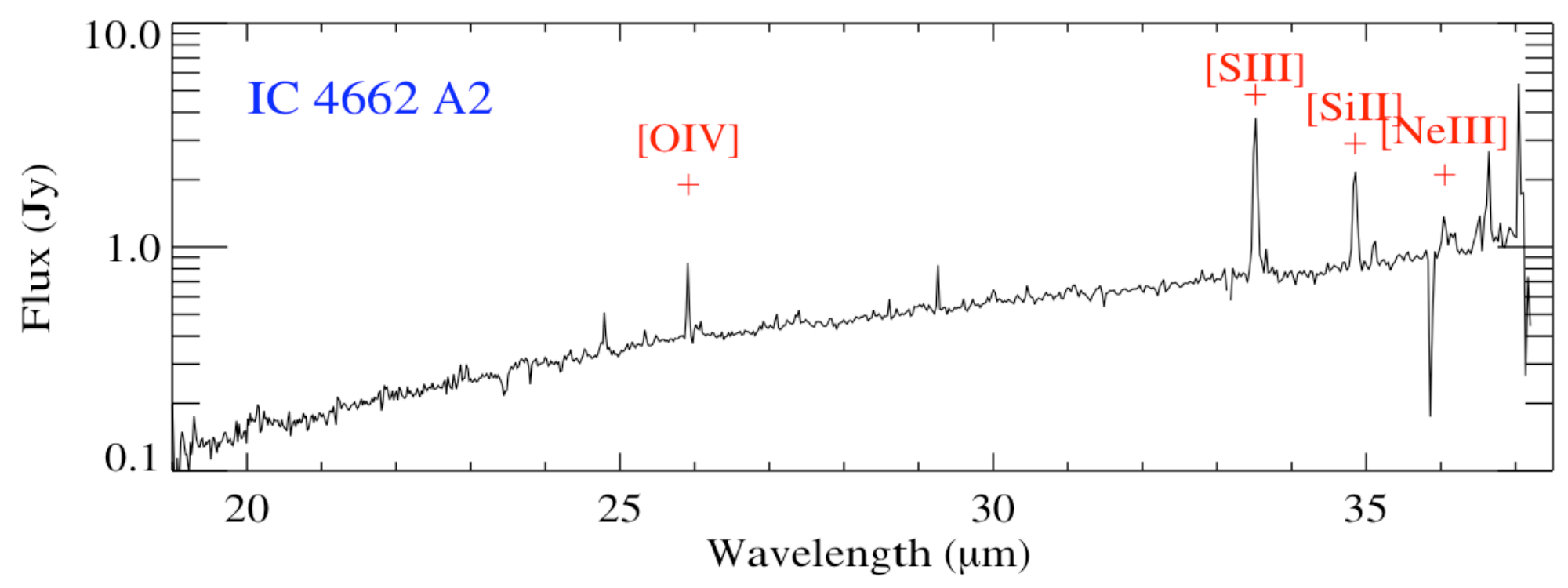}
\caption{IRS LH spectra of IC~4662 regions A1 and A2}
\label{fig:spec_lh} 
\end{figure}

\section{Extinction, Excitation, and Abundances from Spitzer IRS Spectroscopy}
\label{irs}

We observed IC~4662 regions A1 and A2 with Spitzer IRS in both high-resolution ($\sim 500$ km~s$^{-1}$) settings: short-high (SH) covers about $10-20\ \mu$m with a $\sim 5 \times 11"$ slit, and long-high (LH) covers $19-37\ \mu$m with an $\sim 11 \times  22"$ slit.  Both slits are shown in both nod positions for each object in Figure~\ref{fig:slits} on the continuum-subtracted H$\alpha$  and $8\ \mu$m images.  The spectra were reduced and extracted from the full apertures using SPICE, and background-subtracted using the SPOT background calculator.  Figures~\ref{fig:spec_sh} and \ref{fig:spec_lh} show the SH and LH spectra, which feature a warm dust continuum, many high-excitation fine-structure lines, a weak PAH feature at $11.3\ \mu$m but no discernible silicate absorption, and broad H~{\sc i} Humphreys $\alpha$ (Hu$\alpha$) emission at $12.37\ \mu$m.
Fine-structure lines in the IRS spectra make possible the determination of many properties of the gas, dust, and stars within the emitting regions.  

\subsection{Extinction and Electron Density}
\label{dust}
We can estimate the extinction by comparing theoretical  ratios of fine-structure lines that are well-separated in $\lambda$ with the observed values.  In this case we used [Ne~{\sc iii}] 15.6/36 $\mu$m, which does not vary substantially over a wide range of expected electron temperatures and densities (T$_{\rm e}$, n$_{\rm e}$).  We thus assume T$_{\rm e} = 10^4$ K and n$_{\rm e} = 10^2$ cm$^{-3}$ and use  the mid-IR extinction law for the Small Magellanic Cloud of \citet{weingartner01} to derive  for A1 (A2) an extinction A$_{\rm V} = 11\ (9)$ mag for a foreground dust screen or A$_{\rm V} = 25\ (20)$ mag for dust mixed with the emitting gas.
For either dust geometry, correcting the observed  [S~{\sc iii}] 18.7/33.5 $\mu$m ratio for this extinction leads to n$_{\rm e} \approx 150\ (212) $ cm$^{-3}$ for A1 (A2), so our initial assumption of a low n$_{\rm e}$ gives a consistent result, whereas a higher assumption gives an inconsistent one\footnote{By comparing ratios for [Ne~{\sc iii}]  and [S~{\sc iii}]  we are assuming they are co-located, although their ionization potentials do differ significantly.}.  Our fit to the SED (\S~\ref{seds}) yields a higher n$_{\rm e} \sim 10^3$ cm$^{-3}$, which might be due to aperture differences between SH and LH slits: the larger area viewed by the LH slit could produce larger fluxes for the LH lines and hence drive down  the measured ratio and derived n$_{\rm e}$.  We also note that \citet{hunter01} reported a high {\it gas} density in warm photo-dissociation regions of n $=2.8 \times 10^4$ cm$^{-3}$ from ISO data.   These differing density tracers probe different gas phases that may not share the same emitting volume.  Here we adopt a low value of n$_{\rm e} = 10^2$ cm$^{-3}$ in order to obtain a lower limit on the extinction. 

These values for A$_{\rm V}$ greatly exceed those that were determined based on optical spectroscopy by \citet{heydarimalayeri90}, who reported A$_{\rm V} = 0.27$ mag from Balmer decrement measurements,  or even  \citet{stasinska86}, who reported a much higher optical extinction equivalent to  A$_{\rm V} = 2.5$ mag.  Both groups used  apertures of $4-5"$ FWHM, which is comparable with the spatial extent of the SH aperture, which encompasses the bulk of the mid-IR and H$\alpha$ emission from A1 and A2 (Figure~\ref{fig:slits}), so comparison of our data sets is warranted.  
Only a mixed dust model can reconcile the low optical A$_{\rm V}$ values with the high mid-IR ones.  If the extinction were as low as that inferred from visible wavelengths, the mid-IR line ratios would have the theoretical values.  If the dust were in a foreground screen, the high IR extinction would obscure most of the optical recombination line emission; however, we detect only about three times the optically-derived Lyman continuum flux from mid-IR and radio data, so a screen geometry is ruled out.
In other Spitzer observations of nearby and hence well-resolved massive star-formation regions, dust has been observed to be mixed with the stars and gas \citep[e.g. RCW 49,][]{churchwell04}.

\subsection{Excitation and Massive Star Census}
\label{sec:extinction}

Several of the mid-IR fine-structure line ratios that we can measure from the IRS spectra are useful excitation indicators for star-forming regions.  For example, starbursts and HIIRs form a diagonal band in a plot of [S~{\sc iv}] 10.5/[S~{\sc iii}] 18.7 $\mu$m versus [Ne~{\sc iii}] 15.5/[Ne~{\sc ii}] 12.8 $\mu$m, where higher-excitation objects have higher values of both ratios \citep[e.g.][]{wu06}.  With [S~{\sc iv}] 10.5/[S~{\sc iii}] 18.7 $\mu$m ratios $> 1$ and [Ne~{\sc iii}] 15.5/[Ne~{\sc ii}] 12.8 $\mu$m ratios $>7$,  IC~4662 A1 and A2 both fall in the upper right corner of such a diagram, along with other metal-poor WR systems like NGC~5253 and II~Zw~40 \citep[e.g.][]{verma03}.  \citet{heydarimalayeri90} also report a very high excitation for IC~4662 based on its [Ne~{\sc iii}] 386.9 nm/H$\beta$ ratio, and \citet{mashesse99} report a high effective temperature T$_{\rm eff} \sim 38000$ K from its UV spectrum.

High excitation is caused by the hard radiation field of the most massive stars that are present in SSCs, and we have several handles on the massive star content of IC~4662's star-forming regions from our data.  Assuming that the radio continuum flux is due to free-free radiation from HIIRs as above, we find Q $= 4.6 (3.0) \times 10^{51}$ s$^{-1}$ for A1 (A2), and from the Hu$\alpha$ flux we find Q $= 3.1 (2.6) \times 10^{51}$ s$^{-1}$, in reasonable agreement.  From these Lyman continuum luminosities we infer that the number of hot stars needed to power the HIIRs is $\sim 260-460 $ O7V stars. From e.g. \citet{heydarimalayeri90} it is known that these regions contain WR stars, so using the models of \citet{schaerer98} and Starburst99 \citep{leitherer99} we estimate that $\sim 70-150$ WRs should be present.  This number agrees with that of \citet{heydarimalayeri90}  after correcting for the mid-IR-derived extinction.

\subsection{Abundances}

We calculated neon and sulfur abundances for regions A1 and A2 from the observed fluxes of fine-structure lines relative to that in Hu$\alpha$ at 12.37 $\mu$m, following a procedure similar to that used by \citet{wu07b}, whose IRS abundance study of blue compact dwarf galaxies (BCDs) is useful to compare with ours.   We measured sulfur in two excited states at [S~{\sc iii}] 18.71 $\mu$m and [S~{\sc iv}] 10.51 $\mu$m, which we expect to comprise most of the sulfur at such high excitation.  
Following \citet{wu07b} we assumed that only 10 \% of the total S is in [S~{\sc ii}], since the typical ionic abundance of [S~{\sc ii}] in BCDs is $10-15$ \% of that of [S~{\sc iii}] \citep{izotov94,izotov97}.  For neon we used the fluxes of  [Ne~{\sc ii}] 12.81 $\mu$m and [Ne~{\sc iii}] $15.55\ \mu$m and made no correction for unseen ionization states.  All of these lines were observed with the same slit (SH) as Hu$\alpha$ so that no aperture corrections are needed, as would be needed if we used the LH detections of  [S~{\sc iii}] 33.48 $\mu$m and  [Ne~{\sc iii}] 36 $\mu$m.  

The observed fluxes were corrected for extinction using the mixed model results discussed in \S~\ref{sec:extinction}, and the resulting abundances for A1 (A2) are: Ne/H =  $1.4 \times 10^{-5}$  ($2.0 \times 10^{-5}$)  and S/H = $2 \times 10^{-6}$ ($2.4 \times 10^{-6}$).   In order to compare directly with the BCDs of \citet{wu07b}, we adopt their choices for solar abundances of (Ne/H)$_\odot = 1.2 \times 10^{-4}$ \citep{feldman03} and  (S/H)$_\odot = 1.4 \times 10^{-5}$ \citep{asplund05}.  Thus in solar units we have (Ne/H) $= 0.12 (0.17)$ (Ne/H)$_\odot$ and (S/H) $=  0.14 (0.17) $ (S/H)$_\odot$.   \citet{heydarimalayeri90} report a mean Ne/H abundance of $2.25 \times 10^{-5}$, which is consistent with our lower limits but nearly a factor of two larger. 
We note that the  Hu$\alpha$ is significantly broader than the other nebular lines in the IRS spectra.
We investigated whether this broadening could be the direct signature of WR stars via their   He~{\sc i}/{\sc ii} 12.36 $\mu$m emission \citep{jdsmith01}.  However, the number of WR stars needed to explain the putative excess Hu$\alpha$ flux is far larger than the number of O stars expected to be present.  Therefore the cause of the broad Hu$\alpha$ line remains uncertain.

The relatively weak PAH feature at 11.3 $\mu$m is consistent with the trends found by \citet{wu06} for a sample of BCDs that have high excitation and low abundance like  IC~4662 regions A1 and A2.

\section{Conclusions}

The nearby dwarf irregular galaxy IC~4662 harbors two sites of recent massive star formation that have thermal radio spectra, bright dust and H$\alpha$ emission, and rich mid-IR spectra that feature nebular fine-structure lines, a hint of PAHs at 11.3 $\mu$m, and broad Hu$\alpha$ emission.  Our emission-line analysis indicates that this starburst has high excitation and low abundances like many UCD galaxies and in agreement with literature values.  However, we find much higher extinctions (A$_{\rm V} \sim 20-25$ mag) than previous shorter-wavelength studies, which can only be reconciled with the optical observations by a mixed geometry for the gas and dust in these regions.
SED fitting of the radio-to-near-IR SEDs of the UDHIIs in regions A1 and A2 suggests that they are more evolved toward the SSC stage than similar embedded objects (e.g. in He $2-10$ and NGC~5253):  A1 and A2 have lower gas masses in their HIIRs and dust masses surrounding them, and this is consistent with their older ages of about 4 Ma, inferred from radio and optical/near-IR data.  The clusters that power the UDHIIs have masses of about $3 \times 10^5$ M$_\odot$ (assuming a full Kroupa IMF), but they may not be massive, compact, monolithic SSCs like those found in many starbursts, but rather clusters of $10^3-10^4$ M$_\odot$ clusters that fill a larger volume of $~\sim 50$ pc diameter. 

%% Acknowledgements
%
 \acknowledgments
% <Acnowledgments text>
This work was performed in part under the auspices of the U.S. Department of Energy, National Nuclear Security Administration by the University of California, Lawrence Livermore National Laboratory under contract No. W-7405-Eng-48, and it was also supported by The Aerospace Corporation's Independent Research and Development Program.

%\bibliographystyle{spr-mp-nameyear-cnd}  %% BibTeX style
%\bibliography{ssc2}                %% BibTeX data

%%
%% End of file `template.tex'.

\end{document}